\documentclass[aps,twocolumn,superscriptaddress,showpacs,showkeys,amssymb,amsmath,noeprint]{revtex4-2}

\usepackage{dcolumn}

\usepackage{amssymb,amsmath,framed,amsthm}
\usepackage{graphicx}
\usepackage{epstopdf, epsfig}
\usepackage[unicode=true,pdfusetitle,bookmarks=true,bookmarksnumbered=false,bookmarksopen=false,breaklinks=false,pdfborder={0 0 0},backref=false,colorlinks=true]{hyperref}
\hypersetup{citecolor=blue,filecolor=blue,linkcolor=blue,urlcolor=blue}
\usepackage[normalem]{ulem}
\usepackage{xcolor}

\makeatletter
\@ifpackageloaded{stix} 
{
}{                      
  \usepackage{mathrsfs}
  \usepackage{txfonts}          
  \DeclareFontFamily{OML}{txmi1}{}
  \DeclareFontShape{OML}{txmi1}{m}{it}{<->txmi1}{}
  \DeclareSymbolFont{myletter}{OML}{txmi1}{m}{it}
  \DeclareMathSymbol{v}{\mathalpha}{myletter}{`v}

}
\ifx\bm\undefined
  \newcommand{\bm}[1]{{\mathbf{#1}}} 
\else
  \renewcommand{\bm}[1]{{\mathbf{#1}}} 
\fi
\makeatother

\newcommand{\p}{\partial}
\newcommand{\f}[2]{\frac{#1}{#2}}
\newcommand{\mr}[1]{\mathrm{#1}}
\newcommand{\lf}{\left}
\newcommand{\ri}{\right}

\newcommand{\dd}[2]{\frac{\rmd#1}{\rmd#2}}
\newcommand{\pp}[2]{\frac{\p #1}{\p #2}}

\newcommand{\nbl}{\nabla}
\newcommand{\para}{{||}}
\newcommand{\+}{{\perp}}
\newcommand{\unit}[1]{\hat{\bm{#1}}}

\makeatletter
\newcommand{\vast}{\bBigg@{4}}
\newcommand{\Vast}{\bBigg@{5}}

\newcommand{\zhat}{\unit{z}}

\newcommand{\compr}{\mr{compr}}
\newcommand{\AW}{\mr{AW}}
\newcommand{\KAW}{\mr{KAW}}
\makeatother

\def\apjs{Astrophys.~J.~Supp.~Ser.}
\def\apj{Astrophys.~J.}
\def\apjl{Astrophys.~J.~Lett.}

\def\mnras{Mon.~Not.~R.~Astron.~Soc.}

\def\prl{Phys.~Rev.~Lett.}

\DeclareMathAlphabet{\mathpzc}{OT1}{pzc}{m}{it}

\newcommand{\rmc}{\mathrm{c}}
\newcommand{\rmd}{\mathrm{d}}
\newcommand{\rme}{\mathrm{e}}

\newcommand{\rmi}{\mathrm{i}}

\newcommand{\rmA}{\mathrm{A}}
\newcommand{\rmB}{\mathrm{B}}

\begin{document}

\title{Ion versus electron heating in compressively driven astrophysical gyrokinetic turbulence}

\author{Y. Kawazura}
\email{kawazura@tohoku.ac.jp}
\affiliation{Frontier Research Institute for Interdisciplinary Sciences, Tohoku University, 6-3 Aoba, Aramaki, Aoba-ku, Sendai 980-8578 Japan}
\affiliation{Department of Geophysics, Graduate School of Science, Tohoku University, 6-3 Aoba, Aramaki, Aoba-ku, Sendai 980-8578 Japan}
\affiliation{Rudolf Peierls Centre for Theoretical Physics, University of Oxford, Clarendon Laboratory, Parks Road, Oxford OX1 3PU, UK}
\author{A.~A. Schekochihin}
\affiliation{Rudolf Peierls Centre for Theoretical Physics, University of Oxford, Clarendon Laboratory, Parks Road, Oxford OX1 3PU, UK}
\affiliation{Merton College, Oxford OX1 4JD, UK}
\author{M. Barnes}
\affiliation{Rudolf Peierls Centre for Theoretical Physics, University of Oxford, Clarendon Laboratory, Parks Road, Oxford OX1 3PU, UK}
\author{J.~M. TenBarge}
\affiliation{Department of Astrophysical Sciences, Princeton University, Princeton, NJ 08544, USA}
\author{\\Y. Tong}
\affiliation{Space Sciences Laboratory, University of California, Berkeley, CA 94720, USA}
\author{K.~G. Klein}
\affiliation{Lunar and Planetary Laboratory, University of Arizona, Tucson, AZ 85719, USA}
\author{W. Dorland}
\affiliation{Department of Physics, University of Maryland, College Park, MD 20742-3511, USA}

\date{\today}

\begin{abstract}
  The partition of irreversible heating between ions and electrons in compressively driven (but subsonic) collisionless turbulence is investigated by means of nonlinear hybrid gyrokinetic simulations.
We derive a prescription for the ion-to-electron heating ratio $Q_\rmi/Q_\rme$ as a function of the compressive-to-Alfv\'enic driving power ratio $P_\compr/P_\AW$, of the ratio of ion thermal pressure to magnetic pressure $\beta_\rmi$, and of the ratio of ion-to-electron background temperatures $T_\rmi/T_\rme$.
It is shown that $Q_\rmi/Q_\rme$ is an increasing function of $P_\compr/P_\AW$.
When the compressive driving is sufficiently large, $Q_\rmi/Q_\rme$ approaches $\simeq P_\compr/P_\AW$.
This indicates that, in turbulence with large compressive fluctuations, the partition of heating is decided at the injection scales, rather than at kinetic scales. 
Analysis of phase-space spectra shows that the energy transfer from inertial-range compressive fluctuations to sub-Larmor-scale kinetic Alfv\'en waves is absent for both low and high $\beta_\rmi$, meaning that the compressive driving is directly connected to the ion entropy fluctuations, which are converted into ion thermal energy. 
This result suggests that preferential electron heating is a very special case requiring low $\beta_\rmi$ and no, or weak, compressive driving.
Our heating prescription has wide-ranging applications, including to the solar wind and to hot accretion disks such as M87 and Sgr A*. 
\end{abstract}

\maketitle


\section{Introduction}
Most astrophysical systems, e.g., the solar wind, low-luminosity accretion disks, supernova remnants, and the intracluster medium, are in a collisionless turbulent state.
The turbulent fluctuations are generally driven by a large-scale free-energy source that is specific to each system.
These fluctuations are cascaded to small scales via nonlinear interactions, and they are converted ultimately into thermal energy.
This process is called turbulent heating.
In a collisionless plasma, heat is generally deposited into ions and electrons unequally, resulting in a two-temperature state, e.g., in the solar wind~\cite{Cranmer2009}, accretion disks around black holes~\cite{Ichimaru1977,Quataert1999}, and the intracluster medium~\cite{Takizawa1999}.
The partition of turbulent energy between ions and electrons is key to understanding many astrophysical phenomena.
Particularly, in the context of accretion disks around black holes, determining the ion-to-electron heating ratio $Q_\rmi/Q_\rme$ is critical for interpreting radio images from the Event Horizon Telescope (EHT)~\cite{EHT2019a}.
While a recent EHT observation was reproduced numerically using general-relativistic magnetohydrodynamic (GRMHD) simulations~\cite{EHT2019e}, the results strongly depend on the $Q_\rmi/Q_\rme$ prescription used (see~\cite{Ressler2015,Ressler2017,Chael2018,Chael2019} for the GRMHD simulations with different models of $Q_\rmi/Q_\rme$).
Thus, a physical determination of $Q_\rmi/Q_\rme$ is required.

Kinetic, rather than fluid, models must be used in order to calculate correctly the heating rates in a weakly collisional plasma.
For the last few years, turbulent heating has been studied by means of particle-in-cell~\citep{Wu2013,Gary2016,Matthaeus2016,Parashar2018,Zhdankin2019,Parashar2019,Arzamasskiy2019,Zhdankin2020b} and gyrokinetic (GK)~\citep{Howes2011,Told2015,BanonNavarro2016,Kawazura2019} simulations.
In these kinetic simulations, turbulence is excited by injection of artificially configured box-scale fluctuations.
Such box-scale fluctuations are intended to mimic the fluctuations that cascade from larger scales. 
In most of the kinetic simulations referenced above~\footnote{With the exception of those of the freely decaying whistler turbulence~\cite{Gary2016}}, the box-scale fluctuations were Alfv\'enic, meaning that the inertial-range turbulence was assumed to be predominantly Alfv\'enic. 
Spacecraft measurements of the solar wind are qualitatively consistent with this assumption, with less than ten percent of the power contained in compressive (slow-mode-like) fluctuations in the inertial range~\cite{Chen2016,Chen2020}.
However, there is no guarantee that inertial-range fluctuations of turbulence in other astrophysical systems are predominantly Alfv\'enic.
For example, in our recent study of turbulence driven by the toroidal magnetorotational instability, we found that the Alfv\'enic and compressive fluctuations are nearly equipartitioned~\cite{Kawazura2020}.

In this paper, we employ nonlinear GK~\cite{Rutherford1968,Howes2006} simulations to calculate $Q_\rmi/Q_\rme$ in collisionless, subsonic turbulence driven by a mixture of externally injected compressive and Alfv\'enic fluctuations.
The GK theory shows that the energy partition between ions and electrons is decided around the ion Larmor scale~\cite{Schekochihin2009}, meaning that the energy that is not destined for the ion heating at the ion Larmor scale will be channeled into the electron heating.
Therefore, the amount of electron heating can be computed as $Q_\rme = P_\AW + P_\compr - Q_\rmi$, where $P_\AW$ and $P_\compr$ are the Alfv\'enic and compressive energy injection.
Since the electron kinetic effects are not necessary to obtain the electron heating, we use a hybrid-GK model in which electrons are treated as a massless, isothermal fluid~\cite{Schekochihin2009}.

To drive the compressive component of the turbulent cascade, we use slow-mode-like fluctuations.
In our previous, purely Alfv\'enically driven GK simulations~\citep{Kawazura2019}, we determined the dependence of $Q_\rmi/Q_\rme$ on the ratio of the ion thermal pressure to the magnetic pressure, $\beta_\rmi = 8\pi n_\rmi T_\rmi/B_0^2$, and on the ion-to-electron temperature ratio $T_\rmi/T_\rme$. 
We found that $Q_\rmi/Q_\rme$ was an increasing function of $\beta_\rmi$, while the dependence on $T_\rmi/T_\rme$ was weak (similar to the result arising from linear analysis of Landau/Barnes damping~\cite{Quataert1999,Howes2010}).
In this work, we determine the dependence of $Q_\rmi/Q_\rme$ on the ratio of the compressive to Alfv\'enic injection power $P_\compr/P_\AW$.
We also investigate the properties of the phase-space spectra to understand the heating mechanisms related to the compressive cascade.

\section{Hybrid Gyrokinetic Model}
We solve a hybrid-GK model in which ions are gyrokinetic, while electrons are treated as a massless, isothermal fluid~\cite{Schekochihin2009}:
\begin{multline}
  \pp{h_\rmi}{t} + v_\para\pp{h_\rmi}{z} + \f{c}{B_0} \lf\{ \langle \chi \rangle_\bm{R} ,h_\rmi \ri\} = \f{Ze}{T_\rmi}\pp{\lf< \chi \ri>_\bm{R}}{t}F_\rmi \\
  + \lf< C[h_\rmi] \ri>_\bm{R} + \f{v_\para \langle a_\mr{ext}\rangle_\bm{R}}{v_\mr{thi}^2}F_\rmi,
  \label{e:h}
\end{multline}
\begin{equation}
  \dd{}{t}\lf( \f{\delta n_\rme}{n_\rme} - \f{\delta B_\|}{B_0} \ri) + \nbl_\| u_{\|\rme} + \f{cT_\rme}{eB_0}\lf\{\f{\delta n_\rme}{n_\rme},\, \f{\delta B_\|}{B_0}\ri\} = 0,
  \label{e:delta ne}
\end{equation}
\begin{equation}
  \pp{A_\|}{t} + \nbl_\|\lf( \phi - \f{T_\rme}{e}\f{\delta n_\rme}{n_\rme} \ri) = 0.
  \label{e:A|| evo}
\end{equation}
The electromagnetic fields are determined via the quasineutrality condition and the (parallel and perpendicular) Amp\`ere's law:
\begin{equation}
  \f{\delta n_\rme}{n_\rme} = -\f{Ze\phi}{T_\rmi} + \f{1}{n_\rmi}\int\rmd^3\bm{v}\, \lf< h_\rmi \ri>_\bm{r},
  \label{e:quasineutrality}
\end{equation}
\begin{equation}
  u_{\para\rme} = \f{c}{4\pi Zen_\rmi}\nbl_\+^2A_\| + \f{1}{Zen_\rmi}j_\mr{\|ext} + \f{1}{n_\rmi}\int\rmd^3\bm{v}\, v_\| \langle h_\rmi \rangle_\bm{r},
  \label{e:para Ampere}
\end{equation}
\begin{equation}
  \f{B_0}{4\pi}\nbl_\+ \delta B_\| = -n_\rme T_\rme\nbl_\+\lf( \f{\delta n_\rme}{n_\rme} - \f{e\phi}{T_\rme} \ri) + \f{ZeB_0}{c}\int\rmd^3\bm{v}\,\langle (\zhat\times\bm{v}_\+)h_\rmi \rangle_\bm{r},
  \label{e:perp Ampere}
\end{equation}
where $e$ is the elementary charge, $Ze$ is the ion charge, $c$ is the speed of light, $\bm{B}_0$ is the ambient magnetic field, $z$ is the coordinate along $\bm{B}_0$, $(x, y)$ is the plane perpendicular to $\bm{B}_0$, $\bm{v}$ is the particle velocity, $F_\rmi$ is the ion equilibrium distribution function, assumed to be Maxwellian, $\delta f_\rmi = h_\rmi - Ze\phi/T_\rmi$ is the perturbed ion distribution function, $n_\rmi$ and $T_\rmi = m_\rmi v_\mr{thi}^2/2$ are the ion density and temperature associated with $F_\rmi$, $\chi = \phi - \bm{v}\cdot\bm{A}/c$ is the GK potential, $C[\dots]$ is the Coulomb collision operator, $\lf< \dots \ri>_\bm{R}$ is the gyroaverage at fixed gyrocenter position $\bm{R}$, $\lf< \dots \ri>_\bm{r}$ is the gyroaverage at fixed particle position $\bm{r}$, $\delta n_\rme$ is the electron density perturbation, $u_{\|\rme}$ is the parallel electron flow velocity, $n_\rme$ and $T_\rme$ are electron equilibrium density and temperature, $\phi$ is the perturbed electrostatic potential, $A_\|$ is the parallel component of the perturbed vector potential, $\rmd/\rmd t = \p_t + (c/B_0)\{ \phi,\, \dots \}$, $\nbl_\| = \p_z - (1/B_0)\{ A_\|,\, \dots \}$, and $\{ f,\, g \} = (\p_xf)(\p_yg) - (\p_xg)(\p_yf)$.
The remaining symbols follow standard notation.
The compressive fluctuations are driven by an external parallel acceleration $a_\mr{ext}$ in the ion-GK equation~\eqref{e:h}~\cite{Schekochihin2019}, while the Alfv\'enic fluctuations are driven by an external current $j_\mr{\| ext}$ in the parallel Amp\`ere's law~\eqref{e:para Ampere}~\cite{Howes2008a,Howes2011,TenBarge2014,Told2015,BanonNavarro2016,Kawazura2019}.
We consider an electron-proton plasma ($Z = 1$).

The energy budget of the hybrid-GK system is 
\begin{equation}
  \dd{W_\mr{tot}}{t} = P_\AW + P_\compr - Q_\rmi - Q_\rme,
\end{equation}
where
\begin{equation}
  W_\mr{tot} = \int\rmd^3\bm{r}\lf[ \int\rmd^3\bm{v}\, \f{T_\rmi\delta f_\rmi^2}{2F_\rmi} + \f{n_\rme T_\rme}{2}\lf( \f{\delta n_\rme}{n_\rme} \ri)^2 + \f{|\delta \bm{B}|^2}{8\pi} \ri]
  \label{e:W}
\end{equation}
is the free energy,
\begin{equation}
  P_\AW = \int\rmd^3\bm{r}\,\f{j_\mr{\|ext}}{c}\pp{A_{\para}}{t} 
  \label{e:P_AW}
\end{equation}
is the Alfv\'enic injection power, 
\begin{equation}
  P_\compr = \int\rmd^3\bm{r}\int\rmd^3\bm{v}\, \f{T_\rmi a_\mr{ext} v_\| \lf< h_\rmi \ri>_\bm{r}}{v_\mr{thi}^2}
  \label{e:P_compr}
\end{equation}
is the compressive injection power, and
\begin{equation}
  Q_\rmi = -\int\rmd^3\bm{v}\int\rmd^3\bm{R}\,\f{T_\rmi h_\rmi \lf< C[h_\rmi] \ri>_\bm{R}}{F_\rmi}
  \label{e:Qi}
\end{equation}
is the ion heating rate~\cite{Schekochihin2019}.
The electron heating rate $Q_\rme$ is calculated via the hyperresistive and hyperviscous dissipation of the isothermal electron fluid, which are added to Eqs.~\eqref{e:delta ne} and \eqref{e:A|| evo}, respectively~\cite{Kawazura2018}.
In a statistically steady state, $P_\AW + P_\compr = Q_\rmi + Q_\rme$, where each term is time averaged.

This hybrid model is valid at $k_\+ \ll \rho_\rme^{-1}$. 
When $k_\+ \ll \rho_\rmi^{-1}$, the system follows the equations of kinetic reduced MHD (RMHD) wherein compressive fluctuations are passively advected by the Alfv\'enic ones (``Alfv\'en waves'',  AW), and the two types of fluctuations are energetically decoupled~\cite{Schekochihin2009,Meyrand2019}.
The free energy~\eqref{e:W}, therefore,  can be split as $W_\mr{tot} = W_\AW + W_\compr$, where
\begin{align}
  W_\AW =& \int\rmd^3\bm{r} \lf( \f{c^2}{v_\rmA^2}\f{\delta E_\+^2}{8\pi} + \f{\delta B_\+^2}{8\pi} \ri), 
  \label{e:W_AW} \\
  W_\compr =& \int\rmd^3\bm{r} \lf[ \f{n_\rme T_\rme}{2}\lf( \f{\delta n_\rme}{n_\rme} \ri)^2 + \f{\delta B_\|^2}{8\pi} + \int\rmd^3\bm{v}\f{T_\rmi\langle g_\rmi^2 \rangle_\bm{r}}{2F_\rmi} \ri]
  \label{e:W_compr}
\end{align}
where $\delta E_\+$ is the fluctuating perpendicular electric field, $v_\rmA$ is the Alfv\'en speed, and $g_\rmi = \langle \delta f_\rmi \rangle_\bm{R}$.
In the RMHD range, Alfv\'enic fluctuations follow fluid equations, whereas the compressive fluctuations are determined by the ion drift kinetic equation~\cite{Schekochihin2009,Meyrand2019}.
Therefore, in the RMHD range, only ion heating can occur through the phase mixing of compressive fluctuations.  

When $\rho_\rmi^{-1} \ll k_\+ \ll \rho_\rme^{-1}$, the system follows kinetic electron RMHD (ERMHD)~\cite{Schekochihin2009}, which includes two types of fluctuations, ion entropy fluctuations and kinetic AWs (KAWs)
\footnote{
Here and in what follows, we refer, ``colloquially'', to the RMHD- and ERMHD-range turbulent fluctuations as AWs or KAWs. 
This should not be read to imply that we believe turbulence in these regimes to be a collection of random phased weakly interacting waves.
Turbulence in these regimes is critically balanced~\cite{Goldreich1995} and so always strong, and capable of producing intermittent structures, current sheets, etc. 
The reference to ``waves'' simply highlights the fact that even in this strong regime, the linear response relations between fluctuations of different fields, e.g., $\delta E_\+$, $\delta B_\+$, $\delta B_{||}$, $\delta u_{\|\rmi}$, $\delta n_\rme$, etc., are of the same physical nature as in AWs or KAWs (and indeed follow those quite closely even quantitatively).
This is because critical balance implies that linear and nonlinear physics are always of the same order.
A detailed study of this topic can be found in \cite{Groselj2019}.
}.
These fluctuations are again decoupled, and the former are passively advected by the latter.
While the KAWs are ultimately dissipated into electron thermal energy, the ion entropy fluctuations lead to ion heating through phase mixing~\cite{Schekochihin2009}.

There are two types of phase mixing in the GK approximation that cause heating: linear Landau/Barnes damping~\cite{Landau1946,Barnes1966} and nonlinear phase mixing~\cite{Schekochihin2008,Schekochihin2009,Tatsuno2009,Plunk2010}.
The former creates small-scale structure of the distribution function in the $v_\|$ direction of velocity space, which is thermalized via $v_\|$ derivatives in the collision operator $C$. 
The nonlinear phase mixing creates small-scale structure in $v_\+$, and the $v_\+$ derivatives in $C$ cause ion heating.
Previous Alfv\'enic-turbulence simulations showed that ion heating occurs in the ERMHD range exclusively via nonlinear phase mixing for low to modest $\beta_\rmi$~\cite{Told2015,BanonNavarro2016,Kawazura2019,Meyrand2019}, while at high $\beta_\rmi$, there is finite ion heating at $k_\+ \lesssim \rho_\rmi^{-1}$ via linear Landau damping~\cite{Kawazura2019}.
We shall see shortly how this scenario is amended when there is compressive driving.

\subsection{Limitations of the hybrid-GK model}\label{ss:limitations}
Before proceeding to the main results, let us discuss the limitations of our method, and why, despite these limitations, this study is worthwhile.

First, GK ignores large-amplitude and high-frequency fluctuations, resulting in the omission of stochastic heating~\cite{Chandran2010,Hoppock2018} and cyclotron-resonance heating~\cite{Cranmer1999}, respectively.
GK also neglects short-parallel-wavelength fluctuations.
When the driving is at a very large (system-size) scale, the large-scale magnetic field serves as an effective mean field for the fluctuations at the smaller scales, and thus the inertial-range turbulence tends to be anisotropic ($k_\| \ll k_\+$) and small-amplitude ($\delta B/B_0 \ll 1$)~\cite{Kraichnan1965,Howes2008c}. 
The frequencies of sufficiently anisotropic fluctuations are well below the ion cyclotron frequency even at $k_\+\rho_\rmi \sim 1$. 
Therefore, the omission of the large-amplitude, high-frequency, and short-parallel-wavelength fluctuations is a reasonable idealization, and it is reasonable and physically meaningful to ask how energy is partitioned between species in such an idealized turbulence. 

Secondly, we assume the background distribution is an isotropic Maxwellian, but a number of studies have reported that pressure anisotropy can play an important role, e.g., in hot accretion flows~\cite{Sharma2007,Sironi2015a,Sironi2015b,Kunz2016}, in the intracluster medium~\cite{Schekochihin2005}, and in high-beta streams of the solar wind~\cite{Bale2009}.
Furthermore, the linear analysis of GK with an anisotropic background pressure found that $Q_\rmi/Q_\rme$ can differ by as much as an order of magnitude~\cite{Kunz2018} from that obtained via the linear analysis of GK with no background pressure anisotropy~\cite{Howes2010}.
The assumption of a Maxwellian background also imposes the absence of nonthermal particles (e.g., kappa distributions~\cite{Petersen2020} or intermittent beams~\cite{Zhdankin2020a,Zhdankin2020b}).  
While these limitations of our model must be acknowledged, we believe they too represent a reasonable idealization: 
the effect of pressure anisotropy can be significant only for high-$\beta_\rmi$ plasmas, so our results in the low-$\beta_\rmi$ regime should be fairly reliable, whereas at $\beta_\rmi > 1$, we do not push our study to such high values of $\beta_\rmi$ as to render it inexcusably suspect.

Thirdly, the use of the hybrid GK approximation implies neglect of the electron Landau damping.
However, in asymptotic terms, energy partition between ions and electrons is determined at the ion Larmor scale~\cite{Schekochihin2009}, and inclusion of electron Landau damping would be an attempt to account for finite-mass-ratio effects.   
The comparison between the previous hybrid-GK~\cite{Kawazura2019} and full-GK simulations~\cite{Told2015} does not suggest that this makes a significant difference --- and, generally speaking, the experience of working to the lowest order in the mass-ratio expansion is that asymptotic theory does better than one might have pessimistically assumed in capturing the fundamental physics of the ion-scale transition.  

Lastly, relativistic effects are neglected in our model.
This may be problematic when applying our results to ultra-high-energy astrophysical systems;
however, we expect that our results are reasonably useful for hot accretion disks, including the central region where the plasma is only transrelativistic (electrons are relativistic while ions are not).
Indeed, recent two-temperature GRMHD simulations of Sgr A* show that the electron temperature there is at most $k_\rmB T_\rme/m_\rme c^2 \sim 10$~\cite{Chael2018}, so the increase in the electron inertia is not large enough to break the scale separation between $\rho_\rmi$ and $\rho_\rme$, which is the main physical characteristic of our plasma that sets its behavior in what concerns energy partition. 
In this sense, ignoring relativity is akin to ignoring finite-mass-ratio effects.
It is worth noting here that, while one might worry that our model is not ``relativistic enough'', currently feasible simulations of relativistic kinetic turbulence tend to be ``too relativistic'', meaning that they employ the effective mass ratio smaller than the realistic value~\cite{Zhdankin2019,Zhdankin2020b}. 
Here, as is the case of our neglect of electron Landau damping, we choose to err on the side of greater asymptoticity.

\section{Numerical setup}\label{s:numerical setup}
We solve the hybrid-GK model using the \texttt{AstroGK} code~\cite{Numata2010,Kawazura2018} with two sizes of the simulation domain: the ``fiducial'' box $0.125 \le k_x\rho_\rmi, k_y\rho_\rmi \le 5.25$ and the ``double-sized'' box~$0.0625 \le k_x\rho_\rmi, k_y\rho_\rmi \le 5.25$. 
The grid resolution of the phase-space is $(n_x, n_y, n_z, n_\lambda, n_\varepsilon) = (128, 128, 32, 32, 16)$ for the fiducial box, where $\lambda = v_\+^2/v^2$ is the pitch angle, and $\varepsilon = v^2/2$ is the particle's kinetic energy.
Although this grid resolution is lower than that used in some studies~(e.g., \cite{Told2015,Franci2018,Cerri2018}), this is the price for being able to carry out an adequately broad parameter scan.
We will also simulate a single higher-velocity-space-resolution case $(n_x, n_y, n_z, n_\lambda, n_\varepsilon) = (128, 128, 32, 64, 32)$ to check the numerical convergence. 
A recursive expansion procedure~\cite{Howes2008b} is employed to reduce the numerical cost of achieving a statistically steady state.

An oscillating Langevin antenna~\cite{TenBarge2014} is employed to drive the Alfv\'enic and compressive fluctuations.
We choose $(k_x/k_{x0}, k_y/k_{y0}, k_z/k_{z0}) = (1, 0, \pm 1)$ and $(0, 1, \pm 1)$ for the driving modes ($\bm{k}_0$ is the box-size wave number), $0.9\omega_\mr{A0}$ for the driving frequency ($\omega_\mr{A0}$ is the box-size Alfv\'en frequency), and $0.6\omega_\mr{A0}$ for the decorrelation rate.
The amplitude of the Alfv\'en antenna and, therefore, the power of the Alfv\'enic driving, $P_\AW$, is tuned so that critical balance~\cite{Goldreich1995} holds at the box scale~\cite{TenBarge2014}.
We set the same frequency for the compressive driving and Alfv\'enic driving because the compressive fluctuations are passively advected by AWs in the RMHD range.

The ion entropy fluctuations are dissipated by the ion collision operator $C$.
In our code, we employ a fully conservative linearized collision operator~\cite{Abel2008,Barnes2009} and set the collision frequency to $0.005\omega_\mr{A0}$, meaning that ions are almost collisionless.
Since the spatial resolution of our simulation is not sufficient to dissipate all of the ion entropy fluctuations via collisions, we add to $C$ a hypercollisionality term proportional to $k_\+^8$.
Its contribution to ion heating is added to Eq.~\eqref{e:Qi}.
For the dissipation of KAWs, we employ hyperresistivity and hyperviscosity terms proportional to $k_\perp^8$~\cite{Kawazura2018} in the isothermal electron fluid~\eqref{e:delta ne} and \eqref{e:A|| evo}.

Given this setup, the free parameters are $\beta_\rmi$, $T_\rmi/T_\rme$, and the relative amplitude of the compressive driving, which sets $P_\compr/P_\AW$. 
We investigate $\beta_\rmi = (0.1, 1, 4)$ and $T_\rmi/T_\rme = (1, 10)$.
For each case, we consider a range of values of $P_\compr/P_\AW$.

\section{Ion vs. electron heating}
\begin{figure}[b]
  \begin{center}
    \includegraphics*[width=0.49\textwidth]{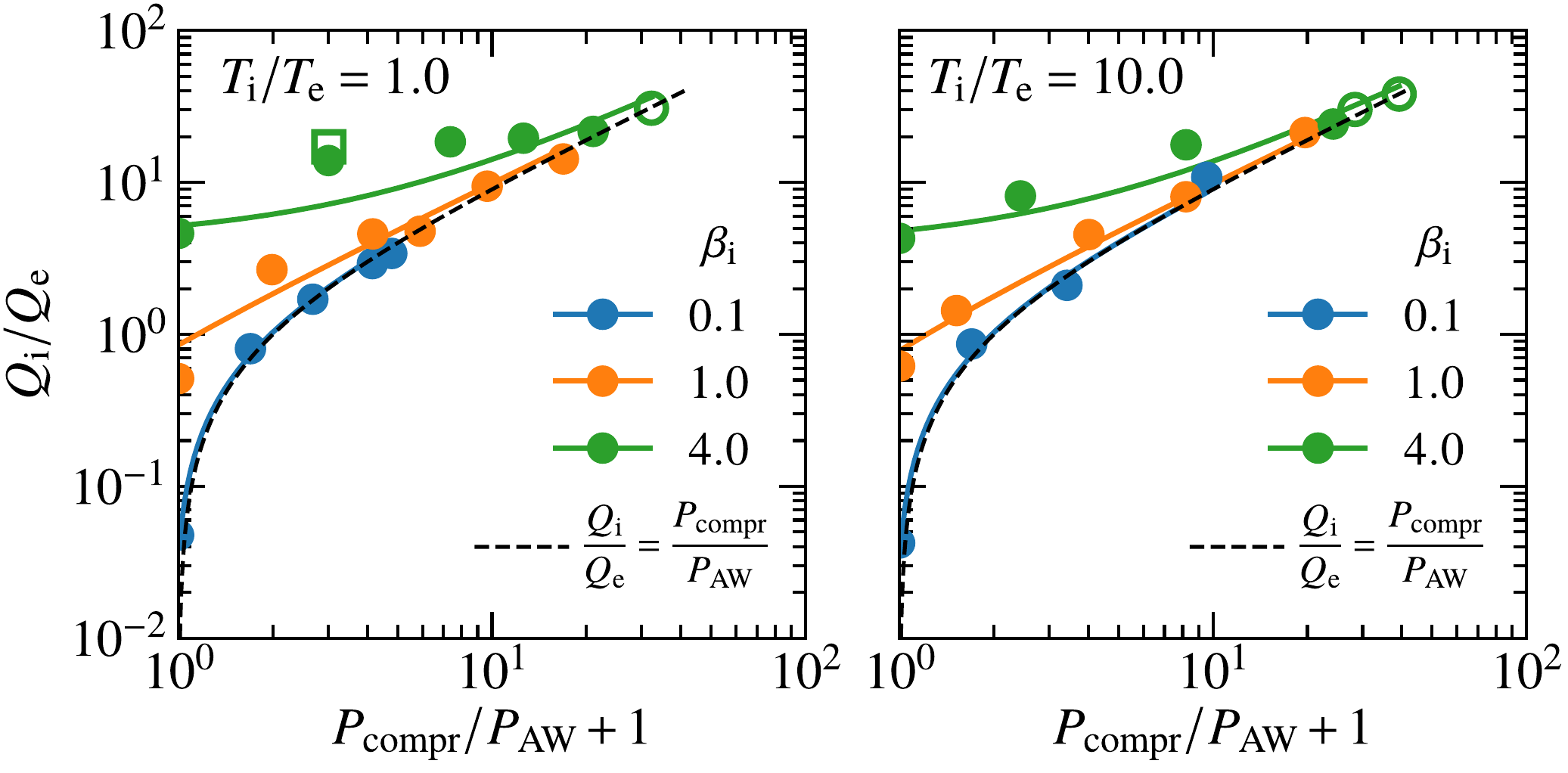}
  \end{center}
  \caption{Dependence of ion-to-electron heating ratio $Q_\rmi/Q_\rme$ on $P_\compr/P_\AW$ for $T_\rmi/T_\rme = 1$ (left) and $T_\rmi/T_\rme = 10$ (right). The markers are simulation results, and the lines are the prescription~\eqref{e:prescription}. The colors correspond to different values of $\beta_\rmi$. The dashed lines correspond to $Q_\rmi/Q_\rme = P_\compr/P_\AW$. The closed circles correspond to the ``fiducial'' box runs, the open circles to the ``double-sized'' box runs, and the green open square to the higher-velocity-space-resolution run (see Sec.~\ref{s:numerical setup}). The value of $Q_\rmi/Q_\rme$ obtained in this higher-velocity-space-resolution run is nearly identical to that of the ``fiducial'' box run with the same ($\beta_\rmi,\, T_\rmi/T_\rme, P_\compr/P_\AW$), demonstrating numerical convergence with respect to the velocity grid.}
  \label{f:QiQe}
\end{figure}
Figure~\ref{f:QiQe} shows the dependence of $Q_\rmi/Q_\rme$ on $P_\compr/P_\AW$ for various values of $(\beta_\rmi,\, T_\rmi/T_\rme)$.
When $P_\compr/P_\AW = 0$, we recover our previous Alfv\'enic results~\cite{Kawazura2019}.
When compressive driving is present, $Q_\rmi/Q_\rme$ is an increasing function of $P_\compr/P_\AW$ for all sets of $(\beta_\rmi,\, T_\rmi/T_\rme)$ that we investigated.
When $\beta_\rmi = 0.1$, $Q_\rmi/Q_\rme = P_\compr/P_\AW$ holds for all $P_\compr/P_\AW$, meaning that all of the compressive power is converted into ion heating, and all Alfv\'enic power is converted into electron heating.
This result was theoretically predicted in~\cite{Schekochihin2019}, and is easy to understand physically: when $\beta_\rmi \ll 1$, ions are too slow to resonate with AWs, and so the Alfv\'enic cascade goes from the RMHD to ERMHD regime without losing power and then gets dissipated on electrons.
What is both new and surprising in our present numerical result is that, even for $\beta_\rmi > 1$,  $Q_\rmi/Q_\rme$ approaches $P_\compr/P_\AW$ when $P_\compr/P_\AW$ is large.
In other words, regardless of $\beta_\rmi$, almost all the compressive fluctuations in the inertial range are converted into ion heat, if the compressive fluctuations are sufficiently large compared to the Alfv\'enic fluctuations.

The comparison of the left and right panels in Fig.~\ref{f:QiQe} suggests that $Q_\rmi/Q_\rme$ does not depend on $T_\rmi/T_\rme$, which already has been seen for the purely Alfv\'enic case~\cite{Kawazura2019};
here we find that it appears to be true also for the compressively driven case.
Admittedly, only two $T_\rmi/T_\rme$ cases ($T_\rmi/T_\rme = $ 1 and 10) have been investigated in our simulation campaign.
Therefore, the weak dependence on $T_\rmi/T_\rme$ that is suggested by the present simulations may cover only the values $T_\rmi/T_\rme \gtrsim 1$.
In contrast, when $T_\rmi/T_\rme \ll 1$ and $\beta_\rmi \ll 1$, namely in the Hall limit~\cite{Schekochihin2019}, there is a theoretical expectation of $Q_\rmi/Q_\rme \to 1$ for any $P_\compr/P_\AW$.
Since this theoretical expectation is inconsistent with Eq.~\eqref{e:prescription}, we need to examine $T_\rmi/T_\rme < 1$ to see whether and when the insensitivity of $Q_\rmi/Q_\rme$ to $ T_\rmi/T_\rme$ breaks. 

\begin{figure*}[t]
  \begin{center}
    \includegraphics*[width=1.0\textwidth]{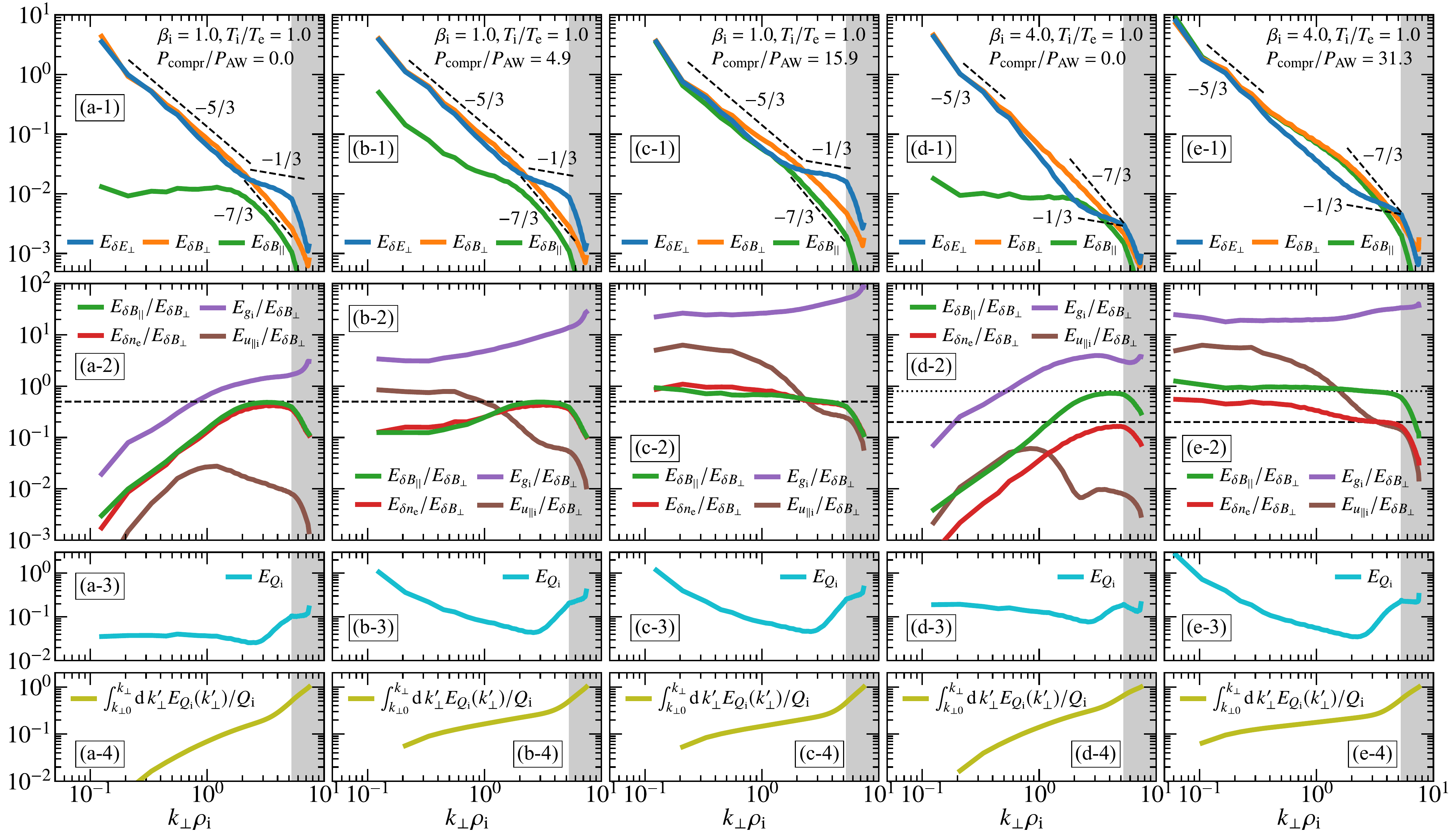}
  \end{center}
  \caption{Top row: power spectra of the electric and magnetic fields, normalized by $\rho_\rmi W_{\delta B_\+}$ where $W_{\delta B_\+}$ is the total perpendicular magnetic energy. Second row: ratios of compressive-field spectra to the perpendicular-magnetic-field spectrum. The horizontal lines correspond to the theoretical predictions for KAWs~\cite{Schekochihin2009} (dotted lines for \eqref{e:Ebpar/Ebper} and dashed lines for \eqref{e:Ene/Ebper}). Third row: spectrum of the ion heating rate normalized by $\rho_\rmi Q_\mr{tot}$ where $Q_\mr{tot} = Q_\rmi + Q_\rme$ is the total heating rate. Bottom row: the ion-heating-rate spectrum integrated up to $k_\+$ and normalized by $Q_\rmi$. Parameter values: (a-1)-(c-4) $\beta_\rmi = 1$, (d-1)-(e-4) $\beta_\rmi = 4$, and $P_\compr/P_\AW$ is increased from left to right. The gray shaded region contains the corner modes in the $(k_x, k_y)$ plane. The results shown in (a-1)-(d-4) and (e-1)-(e-4) are from simulations done in the ``fiducial'' and ``double-sized'' boxes, respectively (see Sec.~\ref{s:numerical setup}).}
  \label{f:kspectrum}
\end{figure*}
Summarizing the parameter dependences that we have found, we propose a simple fitting formula for $Q_\rmi/Q_\rme$:
\begin{multline}
  \f{Q_\rmi}{Q_\rme}(\beta_\rmi,\, T_\rmi/T_\rme,\, P_\compr/P_\AW) \\= \f{35}{1 + (\beta_\rmi/15)^{-1.4} \rme^{-0.1/(T_\rmi/T_\rme)}} + \f{P_\compr}{P_\AW}.
  \label{e:prescription}
\end{multline}
The first term is our previous purely-Alfv\'enic formula~\cite{Kawazura2019}.
One finds that $Q_\rmi/Q_\rme \ge 1$, when $P_\compr/P_\AW \ge 1$ for any $\beta_\rmi$ and $T_\rmi/T_\rme$;
the implication is that preferential electron heating occurs only for Alfv\'enic-dominated turbulence at low $\beta_\rmi$.

\section{Power spectra}
In order to investigate the nature of our simulated turbulence, we plot its free energy spectra in the top row of panels of Fig.~\ref{f:kspectrum}.
The energy spectrum of each integrand in Eqs.~\eqref{e:W_AW} and \eqref{e:W_compr} is denoted by $E$ with a corresponding subscript.
We start by looking at the case of purely Alfv\'enic driving ($P_\compr/P_\AW = 0$). 
As expected, the compressive field, $\delta B_\|$, is negligible compared to the Alfv\'enic fields, $\delta B_\+$ and $\delta E_\+$, in the RMHD range.
Alfv\'enic and compressive fluctuations merge at $k_\+\rho_\rmi \sim 1$ and are reorganized into KAWs and ion entropy fluctuations in the sub-$\rho_\rmi$ range. 
In the RMHD range, the spectra of AW turbulence are $E_{\delta B_\+} \sim E_{\delta E_\+} \sim k_\+^{-5/3}$, while in the sub-$\rho_\rmi$ range, the spectra are $E_{\delta B_\+} \sim k_\+^{-7/3}$ and $E_{\delta E_\+} \sim k_\+^{-1/3}$, which match the standard predictions for KAW turbulence~\cite{Schekochihin2009}.
We are not primarily interested in the accuracy of the spectral slopes because the dynamic ranges of either AW or KAW cascades in our simulations are not wide, so these results are not to be viewed as a contribution to the -5/3 vs. -3/2~\cite{Boldyrev2006} or the -7/3 vs. -8/3~\cite{Boldyrev2012} debates.

The panels in the second row of Fig.~\ref{f:kspectrum} show the spectral ratios: $E_{\delta B_\|}$, $E_{\delta n_\rme}$, $E_{g_\rmi}$, and $E_{u_{\|\rmi}}$ divided by $E_{\delta B_\+}$ ($E_{u_{\|\rmi}}$ is the power spectrum of $m_\rmi n_\rmi u_{\|\rmi}^2/2$).
In the ERMHD range, the theoretical predictions based on the linear response for KAW~\cite{Schekochihin2009},
\begin{align}
  \f{E_{\delta B_\|}}{E_{\delta B_\+}} =\, & \f{\beta_\rmi(1 + T_\rme/T_\rmi)}{2 + \beta_\rmi(1 + T_\rme/T_\rmi)},
  \label{e:Ebpar/Ebper}\\
  \f{E_{\delta n_\rme}}{E_{\delta B_\+}} =\, & \f{4}{(1 + T_\rmi/T_\rme)[2 + \beta_\rmi(1 + T_\rme/T_\rmi)]},
  \label{e:Ene/Ebper}
\end{align}
are quite accurately satisfied.
Furthermore, $u_{\|\rmi}$ rapidly drops in the sub-$\rho_\rmi$ range, which is also consistent with the KAW turbulence theory, where $u_{\|\rmi} = 0$~\cite{Schekochihin2009}.
While the transition from AW to KAW turbulence is transparent at $k_\+\rho_\rmi \simeq 1$ for $\beta_\rmi = 1$, the AW scaling starts to break at $k_\+\rho_\rmi \simeq 0.5$, and then KAW scaling starts at $k_\+\rho_\rmi \simeq 2$ for $\beta_\rmi = 4$.
This ``intermediate'' range at high $\beta_\rmi$ was discovered in our previous purely Alfv\'enic $\beta_\rmi = 100$ simulation~\cite{Kawazura2019}. 

Next, we examine how the spectra change when the compressive driving is present.
We start by focusing on the RMHD range.
As the compressive driving increases, the amplitudes of the compressive fields increase.
One finds that the amplitude of $u_{\|\rmi}$ increases more rapidly than those of $\delta B_\|$ and $\delta n_\rme$, and dominates $E_{g_\rmi}$. 
This is because we drive the compressive fluctuations through an external parallel acceleration of ions, $a_\mr{ext}$ [see Eq.~\eqref{e:h}].
The amplitude of $g_\rmi$ is much greater than those of $\delta B_\|$ and $\delta n_\rme$ when the compressive driving is large, meaning that the compressive driving primarily goes to $g_\rmi$ as it includes the contribution from $u_{\|\rmi}$.
On the other hand, examining the top panels of Fig.~\ref{f:kspectrum}, one finds that the Alfv\'enic fields do not change as $P_\compr/P_\AW$ increases, indicating that the compressive driving does not contaminate the Alfv\'enic fields and confirming that the compressive and Alfv\'enic fields are indeed decoupled in the RMHD range.
While this is a theoretical result that has been accepted for some time~\cite{Schekochihin2009}, the theoretical prediction is based on an asymptotic expansion in $k_\+\rho_\rmi \ll 1$ and relies on a number of assumptions --- most importantly, locality of nonlinear interactions, which is not a completely uncontroversial approach (e.g.,~\cite{Eyink2018}).
Thus, the numerical confirmation of the decoupling shown in Fig.~\ref{f:kspectrum} is a nontrivial result, and a confirmation that a certain way of thinking about plasma turbulence problems is a reasonable one, and that reassuringly, asymptotic theory gives one a decent grasp of the problem even when the small parameter --- $k_\+\rho_\rmi$ in this case --- is only moderately small.

Let us now examine the effect of compressive driving on the sub-$\rho_\rmi$-range cascade.
Even with sufficiently large $P_\compr/P_\AW$, the spectra of the KAW fields, $E_{\delta B_\+}$ and $E_{\delta E_\+}$, do not change. 
The absolute values of spectral amplitude are also preserved.
Therefore, the effect of the compressive driving on KAWs is minor.
In contrast, $E_{g_\rmi}$ increases at all scales as $P_\compr/P_\AW$ increases. 
This result means that the compressive fluctuations in the RMHD range are directly connected to the ion entropy fluctuations in the sub-$\rho_\rmi$ range, while the connection with KAWs appears to be absent. 
If there were an energy-transfer path from the inertial-range compressive fluctuations to KAWs, the amplitudes of KAWs in the compressively driven case would be larger than those in the purely Alfv\'enic case because $E_{\delta B_\+}$ and $E_{\delta E_\+}$ are proportional to $\varepsilon_\KAW^{2/3}$, where $\varepsilon_\KAW$ is the energy flux of the KAW cascade~\cite{Schekochihin2009}.
Nonetheless, the comparison of Fig.~\ref{f:kspectrum} (d-1) and (e-1) shows that $E_{\delta B_\+}$ and $E_{\delta E_\+}$ in the compressively driven case are less than double the purely Alfv\'enic ones even for $P_\compr/P_\AW$ larger than 30.  
In the low-$\beta_\rmi$ regime, the absence of a path between the inertial-range compressive fluctuations and KAWs was analytically proven in~\cite{Schekochihin2019}.
Here, even at $\beta_\rmi = 4$, we find that compressive driving affects only the ion-entropy fluctuations. 
This is the reason why $Q_\rmi/Q_\rme \simeq P_\compr/P_\AW$ is satisfied for $P_\compr/P_\AW \gg 1$ even at $\beta_\rmi \gtrsim 1$.

The panels in the third row of Fig.~\ref{f:kspectrum} show the spectrum of the ion heating rate.
For $\beta_\rmi = 1$ and $P_\compr/P_\AW = 0$, most of the ion heating occurs at sub-$\rho_\rmi$ scales.
This heating-rate spectrum is consistent with the full GK simulation at the same parameters, spanning both the ion and electron kinetic scales~\cite{Howes2011,Told2015,BanonNavarro2016}.
As $P_\compr/P_\AW$ increases, the heating rate both in the RMHD range and at sub-$\rho_\rmi$ scales increases.
For $\beta_\rmi = 4$ and $P_\compr/P_\AW = 0$, there is ion heating in the RMHD range with comparable amplitude to the sub-$\rho_\rmi$ heating.
The ion heating in $k_\+\rho_\rmi \lesssim 0.3$ is due to the Landau damping of AWs since there are no compressive fluctuations at $k_\perp\rho_\rmi \lesssim 0.3$.
This indicates that the box scale of $\beta_\rmi = 4$ simulation is not precisely asymptotically in the RMHD range even though the electromagnetic spectra at $k_\perp\rho_\rmi \lesssim 0.3$ look like those of RMHD turbulence, viz., $E_{\delta B_\+} \sim E_{\delta E_\+} \sim k_\+^{-5/3}$.
Similar to the $\beta_\rmi = 1$ case, the heating rate both in the RMHD range and at sub-$\rho_\rmi$ scales increases as $P_\compr/P_\AW$ increases.

We note that ion heating near the injection scale may be an artifact when the compressive driving is present:
recent drift-kinetic simulations~\cite{Meyrand2019} showed that compressive driving directly heated the ions at the injection scale because the turbulent cascade was not yet well developed at that scale.
However, in our simulations, the contribution of the heating at the injection scale to the total heating rate is negligible.
To show this, we plot, in the bottom panels of Fig.~\ref{f:kspectrum}, the ion-heating-rate spectrum integrated up to $k_\+$ and normalized by $Q_\rmi$, viz., $\int_{k_{\+0}}^{k_\+}\rmd k_\+'\, E_{Q_\rmi}(k_\+')/Q_\rmi$, where $k_{\+0}^2 = k_{x0}^2 + k_{y0}^2$.
This is the fraction of ion heating rate contained at the scales larger than $k_\+^{-1}$.
We find for all cases, most of the ion heating ($\sim$80\%) occurs at sub-$\rho_\rmi$ scales.
While the compressive driving increases the heating rate both in the RMHD and sub-$\rho_\rmi$ ranges (the third row of Fig.~\ref{f:kspectrum}), the contribution to the total ion heating is predominantly from the sub-$\rho_\rmi$ range.
It is also evident that the (possibly artificial) box-scale heating in the presence of the compressive driving is negligible, being only $\simeq$5\% of the total.

\section{Velocity-space structure}
\begin{figure*}[htpb]
  \begin{center}
    \includegraphics*[width=1.0\textwidth]{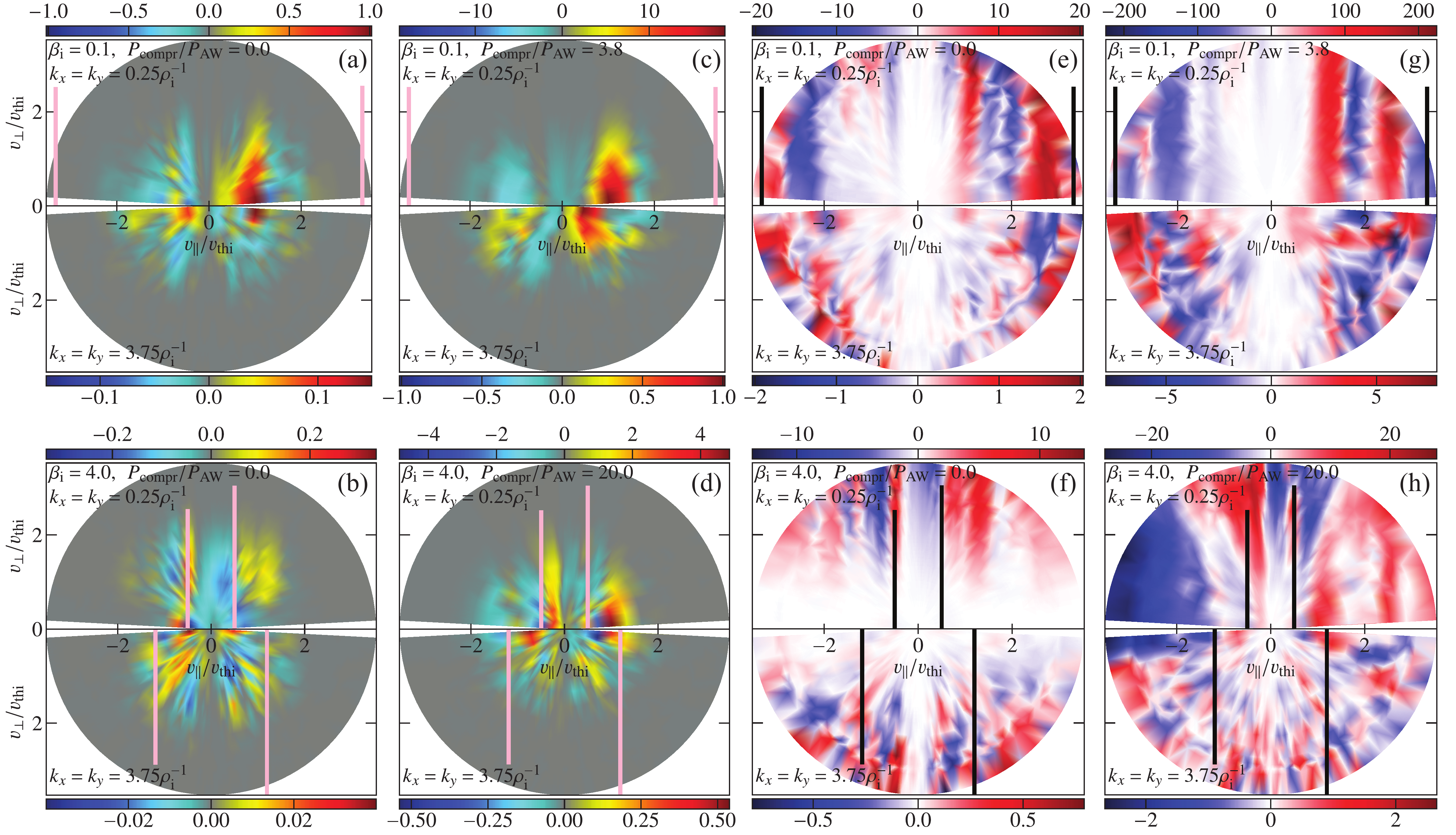}
  \end{center}
  \caption{The real part of the gyroaveraged perturbed ion distribution function $g_\rmi$ (a-d) and $g_\rmi/F_\rmi$ (e-h) in the $z = 0$ plane; $\beta_\rmi = 0.1$ (a, c, e, g) and $\beta_\rmi = 4$ (b, d, f, h); the compressive driving is off (a, b, e, f) and on (c, d, g, h). For each panel, the top half is at $(k_x\rho_\rmi, k_y\rho_\rmi) = (0.25, 0.25)$ and the bottom half is at $(k_x\rho_\rmi, k_y\rho_\rmi) = (3.75, 3.75)$. The vertical pink lines in (a)-(d) and black lines in (e)-(h) correspond to the Landau resonance $v_\| = \pm \omega/k_\| v_\rmA$ which is a solution to the linear dispersion relation of the hybrid-GK model.}
  \label{f:dist_g}
\end{figure*}
In order to investigate the heating process, we show the velocity-space structure of $g_\rmi$. 
We are particularly interested in the small-scale structures of $g_\rmi$ in velocity space as they are the route to heating (i.e., to activating the collision operator) in weakly collisional plasmas~\cite{Krommes1994,Schekochihin2008}.
Figure~\ref{f:dist_g} shows snapshots of $g_\rmi$ and $g_\rmi/F_\rmi$ in the $z = 0$ plane for zero and large compressive driving when $\beta_\rmi = $ 0.1 and 4.
The normalization by $F_\rmi$ helps accentuate the structure at large $|\bm{v}|$~\cite{Li2016}.
In all panels, the top half is taken at $k_x = k_y = 0.375\rho_\rmi^{-1}$, and the bottom half is taken at $k_x = k_y = 5.25\rho_\rmi^{-1}$.
A rough trend is common for both low and high $\beta_\rmi$, with and without the compressive driving:
in the RMHD range, $g_\rmi$ has small-scale structure in the $v_\|$ direction and little structure in the $v_\+$ direction;
in contrast, in the ERMHD range, there is small-scale structure both in $v_\|$ and $v_\+$. 
The small-scale structure in $v_\|$ is due to linear Landau damping~\cite{Landau1946,Watanabe2004,Kanekar2015}; 
the small-scale structure in $v_\+$ is created by nonlinear phase mixing~\cite{Schekochihin2008,Schekochihin2009,Tatsuno2009,Plunk2010}.

\begin{figure}[htpb]
  \begin{center}
    \includegraphics*[width=0.49\textwidth]{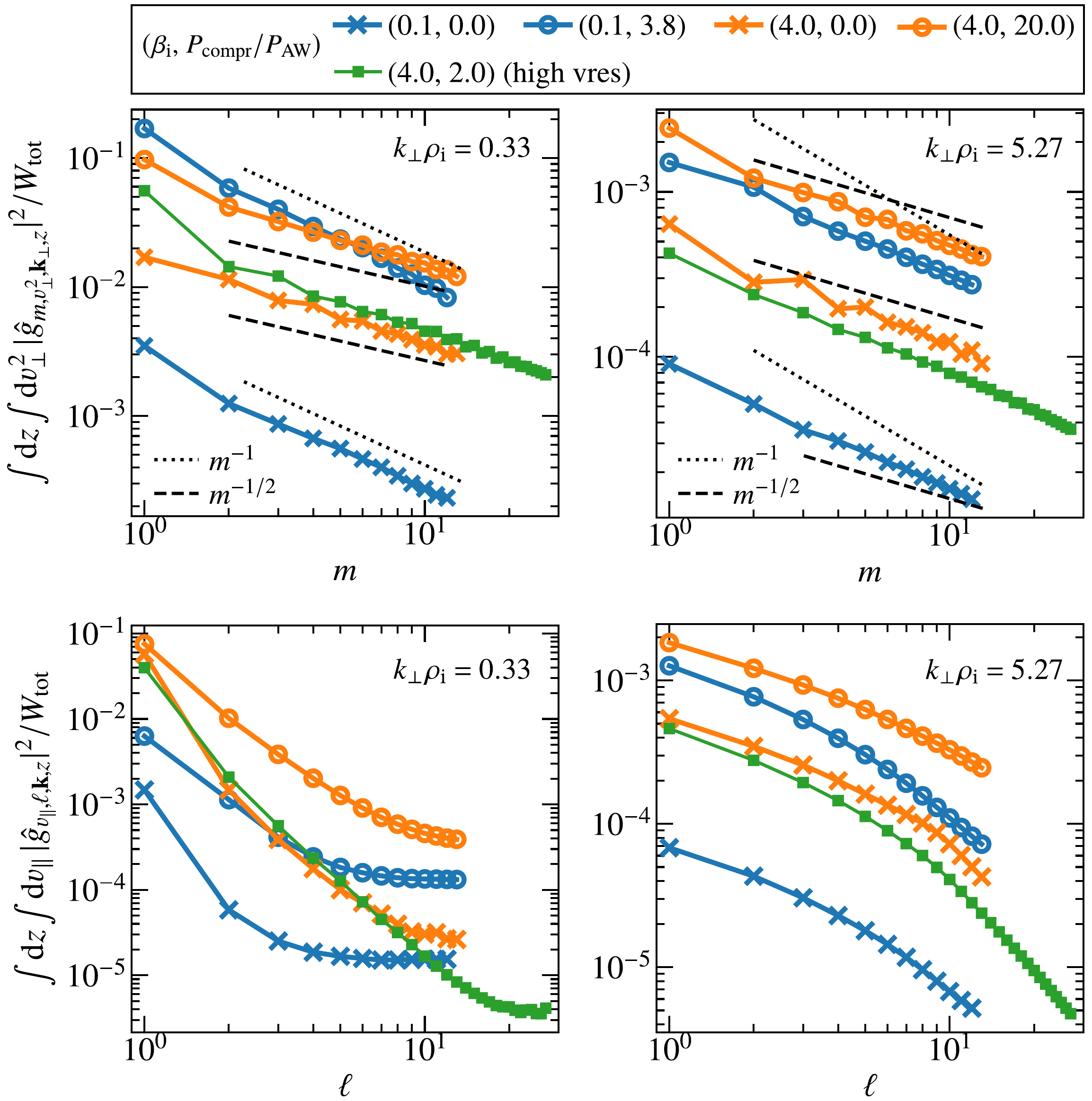}
  \end{center}
  \caption{Hermite (top) and Laguerre (bottom) spectra of the gyroaveraged perturbed ion distribution function $g_\rmi$ (normalized by the total energy) at $k_\+\rho_\rmi = 0.33$ (left) and at $k_\+\rho_\rmi = 5.27$ (right). The blue and orange lines correspond to $\beta_\rmi = 0.1$ and $\beta_\rmi = 4$, respectively. The crosses (circles) correspond to the cases without (with) compressive driving. $m$ and $\ell$ stand for the Hermite and Laguerre moments, respectively. The spectra are integrated over $z$ and $v_\+$ for the Hermite spectrum and over $z$ and $v_\|$ for the Laguerre spectrum. For the Hermite spectra, the auxiliary lines $m^{-1}$ (suggesting suppressed Landau damping/phase mixing~\cite{Schekochihin2016,Adkins2018,Meyrand2019}) and $m^{-1/2}$ (suggesting strong Landau damping~\cite{Schekochihin2016}) are are shown for reference. The green squares correspond to the higher-velocity-space-resolution run, and they are consistent with the ``fiducial'' runs, suggesting numerical convergence with respect to the velocity grid.}
  \label{f:mlspectrum}
\end{figure}

In order to investigate quantitatively the heating mechanism, we examine the Hermite and Laguerre spectra~\cite{Grad1949,Kanekar2015,Schekochihin2016,Servidio2017,Mandell2018,Cerri2018,Adkins2018,Meyrand2019,Kawazura2019} of $g_\rmi$, viz., $|\hat{g}_{m,\ell}|^2$, defined by
\begin{equation}
  \hat{g}_{m,\ell} = \int_{-\infty}^{\infty}\!\rmd v_\para\f{H_m(v_\para/v_\mr{thi})}{\sqrt{2^m m!}}
\int_{0}^{\infty}\!\rmd (v_\perp^2) L_\ell(v_\perp^2/v_\mr{thi}^2)g_\rmi(v_\para, v_\perp^2),
\label{eq:g_def}
\end{equation}
where $H_m(x)$ and $L_\ell(x)$ are Hermite and Laguerre polynomials, respectively. 
The top panels of Fig.~\ref{f:mlspectrum} show the Hermite spectra in the RMHD and ERMHD (sub-$\rho_\rmi$) ranges when the compressive driving is on or off for $\beta_\rmi =$ 0.1 and 4.
The Hermite spectrum quantifies the filamentation in $v_\|$ and indicates whether Landau damping is significant or not: 
the signature of Landau damping is $m^{-1/2}$~\cite{Kanekar2015}; 
a steeper spectrum, which in our simulations is measured to be $m^{-1}$ (cf.~\cite{Schekochihin2016,Adkins2018}), may be an indication that Landau damping (phase mixing) is suppressed by the stochastic echo effect~\cite{Schekochihin2016,Adkins2018,Meyrand2019,Kawazura2019}. 
We find that, at both high and low $\beta_\rmi$, the compressive driving does not change the Hermite spectral slope, viz., $m^{-1}$ for $\beta_\rmi = 0.1$ and $m^{-1/2}$ for $\beta_\rmi = 4$ in the RMHD range and $m^{-1/2}$ both for $\beta_\rmi = 0.1$ and $\beta_\rmi = 4$ in the ERMHD range.
Therefore, regardless of whether the compressive driving exists or not, ion Landau damping is suppressed for $\beta_\rmi = 0.1$ but is active for $\beta_\rmi = 4$ in the RMHD regime
\footnote{
In our previous paper~\cite{Kawazura2019}, we noted that the correspondence between the ion heating and $m^{-1/2}$ spectrum must be viewed cautiously at high $\beta_\rmi$ because the stochastic echo effect may not be computed correctly when the effective collisional cutoff $m_\rmc$ is smaller than $\beta_\rmi$.
In this study, the cutoff is $m_\rmc \sim 10$ while the maximum $\beta_\rmi$ is 4; hence we consider our simulation to be just about safe from this concern
}
\footnote{
We also note that the hypercollisions in our simulations are effective only around the grid scale and do not play any role in the Hermite spectrum at the RMHD range
}
.
Note, however, that the $m^{-1/2}$ spectrum in the ERMHD range should be viewed subject to the following caveat.
Since there is small-scale structure both in $v_\|$ and $v_\+$ directions in ERMHD, and we use $(\lambda,\, \varepsilon)$ grid rather than $(v_\|,\, v_\+)$ grid, the small scale structure in $v_\+$ may contaminate the Hermite spectrum, and thus the $m^{-1/2}$ spectrum may turn out to be a numerical artifact.
Higher velocity-space resolution (currently too expensive) is necessary to determine if this is the case.

The bottom panels of Fig.~\ref{f:mlspectrum} show the Laguerre spectrum, which quantifies the filamentation in $v_\+$ and is, thus, a diagnostic of nonlinear phase mixing.
In contrast to the Hermite spectrum, the Laguerre spectrum in the RMHD range is noticeably modified by compressive driving;
for both $\beta_\rmi$ = 0.1 and 4, the Laguerre spectrum becomes shallower when the compressive driving is present. 
This result indicates that the additional heating in the RMHD range due to compressive driving [Fig.~\ref{f:kspectrum} (b-3), (c-3), and (e-3)] is caused by the emergence of small-scale structures in $v_\+$, presumably triggered by nonlinear phase mixing.
Whereas nonlinear phase mixing has been considered to start at $k_\+\rho_\rmi \sim 1$ in Alfv\'enic turbulence, we find that RMHD-range compressive fluctuations triggers nonlinear phase mixing at $k_\+\rho_\rmi \ll 1$. 
We believe that this is due to the effect of $\nbl \delta B_\|$ drifts~\cite{Howes2006} but leave further investigation of this detail to future work.
In the ERMHD range, on the other hand, compressive driving does not change the Laguerre spectrum.
For both $\beta_\rmi = 0.1$ and 4, the Laguerre spectrum is shallower in the ERMHD range than that in the RMHD range, indicating that the ion heating in the ERMHD range is mediated by the nonlinear phase mixing, as indeed expected theoretically~\cite{Schekochihin2009}.

\section{Conclusions}
In this paper, we have obtained the ion-to-electron irreversible-heating ratio $Q_\rmi/Q_\rme$ in compressively driven (but subsonic) gyrokinetic turbulence.
Summarizing the dependence on the free parameters, $Q_\rmi/Q_\rme$ is (i) an increasing function of $P_\compr/P_\AW$, (ii) an increasing function of $\beta_\rmi$, and (iii) almost independent of $T_\rmi/T_\rme$. 
With regard to (i), $Q_\rmi/Q_\rme \simeq P_\compr/P_\AW$ for any $\beta_\rmi$ when the compressive driving is sufficiently large.
This result suggests that preferential electron heating, $Q_\rmi/Q_\rme \ll 1$, occurs only when $\beta_\rmi \ll 1$ and $P_\compr/P_\AW \ll 1$, a fairly special case. 
A very simple fitting formula for the heating ratio is presented in Eq.~\eqref{e:prescription} and is shown to work remarkably well by Fig.~\ref{f:QiQe}.
This function can be useful in modeling a variety of astrophysical systems, such as the solar wind, AGN jets~\cite{Ohmura2019,Ohmura2020}, and accretion disks around black holes.
Especially for accretion disks, $Q_\rmi/Q_\rme$ is important for interpreting observations by the EHT.
We offer this prescription to the modelers with a word of caution that our results may not be precisely, quantitatively applicable beyond the limitations discussed in Sec.~\ref{ss:limitations}. 
We also note that the parameter sets used for determining our $Q_\rmi/Q_\rme$ function are limited, i.e., $\beta_\rmi = (0.1, 1, 4)$ and $T_\rmi/T_\rme = (1, 10)$.
A wider parameter scan is necessary to extend our prescription Eq.~\eqref{e:prescription} beyond this range, e.g., to the Hall limit, $T_\rmi/T_\rme \ll 1$ and $\beta_\rmi \ll 1$, which may be a special case~\cite{Schekochihin2019}.

We have also analyzed the phase-space spectra of our turbulence to quantify the distribution, and flows, of free energy. 
The spectra show that compressive driving affects the compressive fluctuations in the RMHD range and the ion entropy fluctuations in the sub-$\rho_\rmi$ range, while AWs in the RMHD range and KAWs in the sub-$\rho_\rmi$ range are unaffected. 
This result indicates that compressively injected energy is predominantly converted to ion heating.
The spectra of the ion heating rate (Fig.~\ref{f:kspectrum}) show that most heating happens in the sub-$\rho_\rmi$ range, regardless of whether compressive driving is applied or not.
The analysis of the ion distribution function and its velocity-space spectra quantifies various phase mixing processes, which are routes to free energy thermalization.
We have found that compressive driving does not change the linear phase mixing in the RMHD range, viz., the presence (absence) of phase mixing at high (low) $\beta_\rmi$; however a new channel of heating through the enhanced nonlinear phase mixing in the RMHD range emerges when compressive driving is present. 
While most of these results conform to theoretical expectations~\cite{Schekochihin2009,Schekochihin2016,Schekochihin2019}, ours appears to be the first study in which some of them have received their numerical corroboration.

In order for results like those reported here to be useful in large-scale modelling, the modeler must know how the turbulent energy injected into their plasma system at large (system-size) scales is partitioned into Alfv\'enic and compressive (slow-wave-like) cascades in the inertial range.
This is an unsolved problem in the majority of astrophysical contexts, but it is a solvable one: such a partition is decided at fluid (MHD) rather than kinetic scales.
We hope to present a solution to this problem for turbulence driven by the magnetorotational instability~\cite{Balbus1991} with near-azimuthal mean magnetic field in a forthcoming publication~\cite{Kawazura2020}.

%
%
\begin{acknowledgments}
YK, MAB, and AAS are grateful to S. Balbus and F. Parra for very fruitful discussions.
YK thanks G. Howes for providing the numerical code for the recursive expansion method~\cite{Howes2008b}. 
YK, AAS, and MAB were supported by the STFC grant ST/N000919/1.
YK was supported by JSPS KAKENHI grant JP19K23451 and JP20K14509. 
AAS and MAB were supported in part by the UK EPSRC Grant EP/R034737/1.
JMT was supported by NSF SHINE award (AGS-1622306).
KGK was supported by NASA grant 80NSSC20K0521.
JMT, KGK, and YT acknowledge the 2014 ISSI meeting that originally motivated the development of compressive driving code.
Numerical computations reported here were carried out on the EUROfusion HPC (Marconi--Fusion) under project MULTEI, on ARCHER through the Plasma HEC Consortium EPSRC grant number EP/L000237/1 under projects e281-gs2, on Cray XC50 at Center for Computational Astrophysics in National Astronomical Observatory of Japan, and on the University of Oxford's ARC facility.
\end{acknowledgments}


\bibliography{references}

\end{document}